# AIRKEy: Multimodal Acoustic-Assisted WiFi Sensing for Zero-Training Robust PIN Inference

Baichuan Wu
University of Science and Technology of China
Hefei, Anhui, China
wbc88@mail.ustc.edu.cn

Bin Liu✉
University of Science and Technology of China
Hefei, Anhui, China
flowice@ustc.edu.cn

Xiang Zhang✉
Tianjin University
Tianjin, China
zhangxiang@ieee.org

Zhi Liu
The University of Electro-Communications
Chofu, Japan
liu@ieee.org

Jie Zhang
A*STAR Institute of High Performance Computing
Singapore, Singapore
zhangj6@a-star.edu.sg

Chao Liu
Ocean University of China
Qingdao, Shandong, China
liuchao@ouc.edu.cn

Huan Yan
Guizhou Normal University
Guiyang, Guizhou, China
yh1995.cs@gmail.com

Meng Li
Hefei University of Technology
Hefei, Anhui, China
mengli@hfut.edu.cn

Fusang Zhang
Inspur Computer Technology Co., Ltd.
Jinan, Shandong, China
zhangfusang@buaa.edu.cn

## Abstract

Contactless keystroke inference via WiFi sensing highlights severe privacy threats, yet its real-world feasibility is hindered by two fundamental physical and deployment bottlenecks: the strict requirement for network privileges to acquire stable sensing streams, and the inherent "waveform fusion" ambiguity of pure WiFi signals during rapid, muscle-memory typing. To overcome these limitations, we propose AIRKEy, a novel cross-modal sensing framework that achieves highly stealthy, zero-training PIN eavesdropping. First, to bypass network deployment barriers, AIRKEy exploits fundamental IEEE 802.11 mechanisms to predictably elicit Acknowledgment (ACK) responses from unmodified target devices. By passively harvesting Channel State Information (CSI) from these ACKs using a low-cost microcontroller, AIRKEy secures a continuous spatial sensing stream entirely without network association. Crucially, to resolve the WiFi waveform fusion bottleneck, AIRKEy introduces a cross-modal complementarity mechanism. By utilizing lightweight acoustic signals as precise temporal anchors, the system robustly guides the segmentation of overlapping CSI trajectories. This joint spatiotemporal fusion strictly intersects CSI-derived spatial similarities with acoustic-guided inter-keystroke timing. Extensive real-world evaluations demonstrate that AIRKEy achieves over 4× higher accuracy than state-of-the-art unimodal zero-training schemes, successfully recovering device-unlock PINs within 6 attempts. Ultimately, this work exposes a critical vulnerability in contemporary smart interfaces, underscoring the severe privacy implications of ubiquitous multimodal sensing.

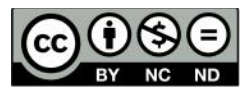

## 1 Introduction

Numerical passwords serve as the primary gatekeepers for digital assets [9, 52]. However, entering these credentials in semi-public spaces exposes users to severe privacy risks from proximate adversaries [23]. While users typically remain vigilant against overt surveillance like cameras [68, 69], covert side-channel leakages persist unnoticed. These invisible channels are covertly exploited by adversaries via physical signals carrying keystroke information, posing a formidable threat that is significantly more difficult to detect and mitigate. Existing keystroke-eavesdropping techniques primarily exploit side channels such as acoustic signals [26, 33, 50, 71], electromagnetic emissions [25], indirect vision [7, 10, 53], and motion sensors [6, 32, 36, 54]. However, their practical efectiveness typically relies on stringent operational assumptions [20]. Notably, while acoustic signals can capture typing rhythms, they inherently lack the fine-grained spatial resolution required to discriminate between diferent keystroke patterns unless utilizing conspicuous, close-proximity (decimeter-level) microphone arrays [21]. Other modalities similarly demand unobstructed line-of-sight [10] or invasive malware [13]. To overcome these constraints, recent research has turned to contactless WiFi sensing [20, 41, 58], which exploits characteristic perturbations in Channel State Information (CSI) or Beamforming Feedback Information (BFI) [51] induced by finger motions. Because WiFi signals can operate over meter-scale distances and blend seamlessly with ambient network traffic, they provide a highly stealthy and practical vector for keystroke inference without requiring direct line-of-sight or software installation.

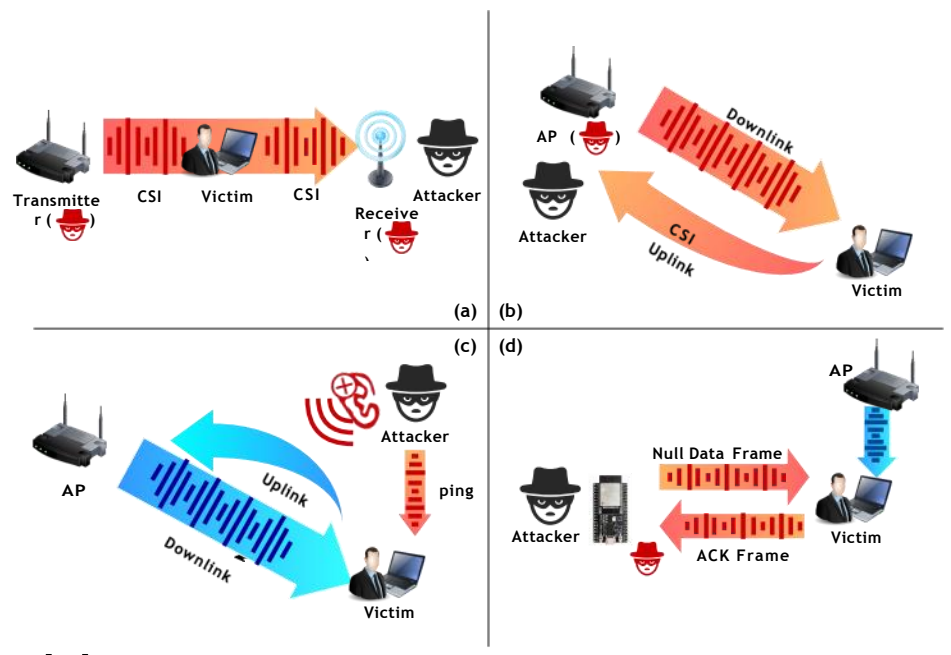

**Figure 1: (a) Attackers deploy conspicuous transceivers; (b) attackers control the victim's AP to access CSI; (c) attackers share the same WiFi network and use ping traffic to obtain stable BFI; (d) AIRKEy use ACK-induced CSI, requiring no shared network or extra transceivers.**

Despite their stealthy nature, current WiFi-based keystroke eavesdropping techniques face significant deployment bottlenecks in real-world environments. The primary limitation stems from the difficulty of acquiring a stable and continuous sensing stream unobtrusively. As illustrated in Figure 1, existing approaches either mandate the deployment of conspicuous WiFi transceivers, requiring the victim to be precisely positioned between them to capture signal perturbations [2, 28, 59, 70], or require control of the access point (AP) to which the victim's device is connected. Specifically, attackers must either control the victim's AP to extract CSI from uplink traffic [29,38], or share the same WiFi network to inject continuous probing packets (e.g., ping traffic) that induce Beamforming Feedback Information (BFI) [9, 20]. Consequently, the practical feasibility of these methods in everyday settings remains highly constrained, particularly when users operate on private networks or remain vigilant against unfamiliar hardware. Under such stringent constraints, numerical passwords are generally assumed to be secure against remote RF sensing. This work revisits that assumption and poses the first fundamental question:

> **Q1:** Can WiFi-based keystroke inference attacks be achieved without any shared networks or additional transceivers?

To answer Q1, we exploit an inherent mechanism of commodity WiFi devices: mandatory Acknowledgment (ACK) responses. Prior work [1] has demonstrated that injecting forged data frames forces target devices to instantly reply with ACKs, bypassing higher-layer validations due to strict IEEE 802.11 SIFS timing [22]. While previous studies have noted the feasibility of extracting CSI from these ACKs for general sensing [1], this behavior has primarily been treated as a network performance anomaly or a Denial-of-Service (DoS) surface. Consequently, vendor mitigations have simply focused on limiting response rates. **Our key insight is that while such rate limits efectively thwart network flooding, the re – sulting CSI stream is entirely sufficient for fine-grained side- channel eavesdropping.** By passively harvesting this induced CSI, an attacker can covertly transform the victim's own device into a continuous sensing signal transmitter, capturing minute signal variations to infer keystrokes without requiring any network privileges. With a stable CSI stream, we adopt a zero-training spatiotemporal inference pipeline [9, 59]. Although deep learning approaches ofer high accuracy, their severe performance degradation across diferent environments and users necessitates impractical data collection and retraining [9, 15, 66]. Conversely, zero-training methods bypass this brittleness by directly exploiting the physical signatures of keystrokes: using CSI waveform similarity to identify repeated keys and inter-keystroke timing to infer spatial distances [59]. However, purely WiFi-based zero-training methods face two fundamental physical bottlenecks in realistic scenarios. First, temporal uncertainty within a single keystroke. Each keystroke waveform comprises a press–contact–release process, and the true keystroke event does not occur at a fixed position within the waveform. As a result, the timing of keystrokes cannot be reliably localized, making accurate estimation of inter-keystroke intervals inherently difficult. Second, *waveform fusion*. During rapid, muscle-memory-driven password entry, WiFi's coarse spatial resolution and severe indoor multipath cause the CSI responses of consecutive keystrokes to overlap. Because the signal perturbation of the first key fails to decay before the second begins, this temporal fusion destroys the spatial distinguishability of individual keys, leading to catastrophic errors. Recognizing that these ambiguities are inherently difficult to resolve using the WiFi modality, we pose the following question:

> **Q2:** How can we overcome the inherent temporal uncertainty within individual keystrokes and the waveform?

To address this challenge, we introduce a cross-modal complementarity mechanism that augments WiFi sensing with lightweight acoustic assistance. Unlike traditional acoustic eavesdropping that requires conspicuous, close-proximity microphone arrays for spatial localization (e.g., via DoA) [21], our approach exploits audio strictly in the temporal domain. Using just two compact microphones equipped with simple beamforming, we capture distinctive keystroke sounds to serve as precise *temporal anchors*. These an - chors not only provide precise temporal features of the keystroke sequence but also rigorously guide the segmentation of the contin- uous CSI stream, facilitating the extraction of more accurate spatial features of the keystroke sequence. Crucially, they empower us to disentangle WiFi waveform fusion: when the acoustic modality detects two distinct keystrokes but the corresponding CSI trace appears as a single merged waveform, this cross-modal mismatch explicitly flags a fusion event. Rather than causing catastrophic seg- mentation errors, this anomaly is translated into a strict spatial con- straint, deducing that the fused signals must originate from spatially adjacent keys or repeated presses, thereby preserving inference robustness. We implement AIRKEy using a highly portable and low- cost setup comprising a commodity WiFi adapter, an ESP32 micro- controller, and two compact microphones. In real-world case stud- ies, without requiring network association or dedicated transceiver infrastructures, AIRKEy successfully disentangles rapid keystrokes and recovers device-unlock PINs within 6 guesses. These results underscore the severe privacy implications of exploiting multimodal sensing in everyday environments. We also observe a general principle: WiFi sensing provides high spatial sensitivity but sufers from temporal ambiguity, while acoustic sensing ofers precise temporal resolution but limited spatial discrimination. Their combination yields a naturally complementary multimodal sensing paradigm. In summary, our main contributions are as follows:

(1) We propose a multimodal sensing framework that leverages acoustic signals to provide temporal cues while exploiting CSI to capture fine-grained spatial features, thereby achieving complementary temporal–spatial sensing and efectively resolving the inherent bottleneck of WiFi and acoustic sensing.
(2) We demonstrate that attackers can passively elicit ACK-induced CSI from unmodified devices without network association or specialized transceivers, significantly broadening the real-world feasibility of WiFi-based eavesdropping.
(3) We build a practical system using commodity hardware and validate its efectiveness in real-world environments, revealing real privacy threats of multimodal WiFi sensing.

## 2 Related Work

**Acoustic Sensing.** Acoustic eavesdropping infers keystrokes by analyzing inter-key timing or sound propagation properties [3, 45]. While highly efective for temporal segmentation, accurate spatial localization inherently requires estimating Direction-of-Arrival (DoA) [21,48,72]. This necessitates either conspicuous microphone arrays placed in close proximity to the victim, or malicious access to the victim's dual-microphone recordings [13, 33]. Consequently, utilizing pure acoustics for fine-grained spatial recognition remains highly constrained by hardware and deployment requirements.

**Vision, Motion, and EM Side Channels.** Previous non-acoustic keystroke eavesdropping techniques have primarily exploited vi- sion, motion sensors, and electromagnetic (EM) emanations. Vision- based methods reconstruct input by tracking finger trajectories [7, 60], device motions [46], or screen reflections [42], but they strictly require unobstructed line-of-sight and are highly sensitive to light- ing conditions [10, 34, 43, 63]. Motion sensor approaches rely on smartphone or smartwatch accelerometers to capture typing vibra- tions [35, 36, 39], while EM techniques monitor human-coupled emanations or power-consumption variations [11, 25, 40, 49]. How- ever, both motion and EM methods typically demand invasive soft- ware access or the deployment of equipment in extreme proximity to the target, severely limiting their real-world practicality.

**WiFi–based Sensing.** WiFi sensing enables ubiquitous, contact - less monitoring using CSI [12, 29, 59, 70] or BFI [8, 9, 20]. Existing approaches generally fall into training-based classifiers [20] and zero-training spatiotemporal inference [9, 59]. Despite their long- range capabilities, current WiFi attacks face critical deployment barriers: they require either conspicuous dedicated transceivers deployed around the victim [59, 70] or strict network privileges (e.g., AP control or network association) to induce stable sensing streams [8, 20, 29]. Furthermore, pure WiFi sensing struggles with spatial ambiguity during rapid typing due to waveform fusion. In contrast, our framework overcomes these single-modality bottle- necks by seamlessly fusing unassociated, ACK-induced CSI with acoustic temporal anchors, eliminating the need for network privi- leges, complex microphone arrays, or extensive training data.

## 3 CSI and Keystroke

In orthogonal frequency-division multiplexing (OFDM) based WiFi communication, the received signal $Y(f, t)$ is the product of the transmitted signal $X(f, t)$ and the CSI $H(f, t)$, with additional noise [30, 61, 64]. This process can be expressed as [37, 55, 56, 65]:

$$Y(f, t) = H(f, t) \cdot X(f, t) + N(t), \tag{1}$$

where $N(t)$ denotes additive Gaussian noise, and $H(f, t)$ char - acterizes the propagation efects experienced by the WiFi signal, including multipath propagation, attenuation, and phase rotation. The per-subcarrier channel response is:

$$H(f) = |H(f)|e^{j\theta(f)}, \tag{2}$$

where $|H(f)|$ and $\theta(f)$ correspond to the signal magnitude (attenuation) and phase (rotation) at frequency $f$, respectively. It is often convenient to decompose $H$ into *static* $H_s$ and *dynamic* $H_d$ components [24, 31, 47, 57, 67]:

$$\begin{aligned} H(f, t) &= H_s(f, t) + H_d(f, t) \\ &= \sum_{m_s \in \Phi_s} a_{m_s}(f, t) e^{-j2\pi \frac{d_{m_s}(t)}{\lambda}} \\ &\quad + \sum_{m_d \in \Phi_d} a_{m_d}(f, t) e^{-j2\pi \frac{d_{m_d}(t)}{\lambda}}, \end{aligned} \tag{3}$$

where $\Phi$ is the set of paths, $\Phi_s$ are static paths (e.g., walls, furniture, static body parts) and $\Phi_d$ are dynamic paths perturbed by motion, such as keystroke. Here, $a_m(f, t)$ and $d_m(t)$ represent the complex attenuation and propagation length of the $m$-th path, $\lambda$ is the wavelength, and $t$ is time.

In our scenario, typing induces centimeter-level micro-motions of the hand within the antenna's near field. These motions perturb dynamic propagation paths ($\Phi_d$ in Eq. 3), generating brief, structured fluctuations in $H(f, t)$ [16]. Because distinct keys occupy unique physical locations, their presses excite diferent multipath clusters. This yields characteristic CSI amplitude and phase evolutions, which serve as fine-grained *spatial signatures* for individual keystrokes. While recent works increasingly adopt BFI [19, 62], BFI is exclusively generated during explicit sounding exchanges (e.g., NDPA/NDP) and is completely absent from standard ACK frames. This inherently prohibits its application in our unassociated sensing paradigm. Conversely, raw CSI can be seamlessly extracted from the ubiquitous ACK responses. Furthermore, rather than relying on re- strictive PC-based network cards (e.g., Intel 5300 or AX210) [14,17], AIRKEy employs a compact, low-cost ESP32 microcontroller [18] to passively harvest these CSI measurements. This plug-and-play module requires only a simple USB connection, enabling highly stealthy and portable data acquisition.

## 4 Methods

### 4.1 System Overview

We consider a proximity-based insider threat where an unassociated attacker infers a victim's PIN using solely of-the-shelf hardware (a commodity WiFi adapter, an ESP32, and dual microphones). Operating strictly without network privileges, AIRKEy achieves zero-training keystroke eavesdropping through a three-stage cross-modal pipeline (Figure 2): **1) Target Identification.** The system passively scans wireless channels and correlates overt physical typing activities with CSI dynamics to definitively isolate the victim's MAC address and operating channel. **2) Cross-Modal Data Acquisition.** To secure a continuous sensing stream, the attacker injects null data frames to seamlessly elicit ACK responses from the target. The ESP32 extracts CSI from these ACKs to capture spatial trajectories, while the microphones concurrently record typing sounds to establish precise temporal anchors. **3) Spatiotemporal PIN Inference.** Leveraging the complementarity between the precise temporal cues provided by acoustic anchors and the fine-grained spatial features captured by CSI trajectories, AIRKEy accurately segments the continuous CSI stream, and efectively resolves the inherent RF waveform fusion bottleneck. Subsequently, the system jointly evaluates the extracted cross-modal spatiotemporal features to significantly constrain the candidate search space, and outputs a possible PIN list via zero-training inference.

## 4.2 Target Identification and Cross-Modal Data Acquisition

The first step in AIRKEy is to pinpoint the victim's MAC address and operating WiFi channel. Using a WiFi network adapter with monitor mode capability, the attacker passively scans the environment to compile a candidate MAC list. This list encompasses both addresses with Organizationally Unique Identifiers (OUIs) matching the victim's device model, and randomized MAC addresses (identifiable via the locally administered bit). To definitively isolate the true target from these candidates, AIRKEy sequentially induces ACKs from each MAC on the list. By visually observing the victim's overt physical movements (e.g., sitting down or shifting posture) and correlating them with simultaneous, pronounced fluctuations in the corresponding CSI stream, the attacker can reliably verify the victim's active MAC address and channel from background traffic.

Once the target is identified, AIRKEy initiates unassociated, multimodal data acquisition. To secure a stable RF sensing stream, we use Scapy [5] to inject standard-compliant null data frames addressed to the victim. By exploiting the strict IEEE 802.11 SIFS timing, these frames predictably elicit immediate ACK responses, bypassing higher-layer validations [1, 22]. A low-cost ESP32 module in monitor mode [18] then passively intercepts these ACKs, extracting 128 bytes of fine-grained CSI (64 subcarriers) from the Legacy Long Training Field (LLTF). Crucially, this induction requires no network privileges and operates at stealthy, low rates (<150 packets/s) with empty payloads, ensuring no noticeable link degradation.

Concurrently, to overcome the inherent waveform fusion bottleneck, AIRKEy captures the acoustic modality to serve as precise temporal anchors. We deploy two compact USB microphones spaced 3.5 cm apart on the attacker's laptop. By calculating the victim's relative azimuth and applying time-diference delay-and-sum beamforming [4], the system efectively suppresses of-axis environmental noise. This synchronized, cross-modal acquisition pipeline ensures that AIRKEy obtains both high-resolution spatial trajectories (via CSI) and accurate temporal segmentation cues (via Audio) for robust zero-training inference.

## 4.3 CSI Preprocessing

To enable robust spatiotemporal feature extraction, we first denoise the raw CSI stream. We apply a 3rd-order zero-phase Butterworth low-pass filter to suppress high-frequency measurement noise without introducing phase distortion. Subsequently, Principal Component Analysis (PCA) is employed to extract the dominant signal features, averaging the first two principal components to form a unified CSI representation. To segment continuous CSI into discrete keystroke events, we leverage two complementary features: waveform energy and dispersion. For the energy-based approach, we apply a sliding window of length $2w + 1$ with a step size of $b$. The local energy $E_i$ ofthe $i$-th window is computed as:

$$E_i = \frac{1}{2w+1} \sum_{k=-w}^{w} |x_{i+k}|^2, \tag{4}$$

where $|x_{i+k}|$ denotes the CSI amplitude. Potential keystrokes are identified as continuous intervals where the energy profile $E_i$ exceeds an adaptive threshold $T_E$. To robustly accommodate environmental variations, we empirically define $T_E = \mu_E + 0.01 * (MAX_E - \mu_E)$, where $\mu_E$ and $MAX_E$ are the mean and maximum of the overall CSI energy. As depicted in Figure 3, this transformation yields prominent peaks that efectively delineate keystroke boundaries from the raw CSI sequence.

To capture abrupt signal fluctuations induced by rapid finger movements, we complement the energy profile with a local dispersion metric. Using the identical sliding window setup, the dispersion $\sigma_i$ is defined as the mean absolute deviation:

$$\sigma_i = \frac{1}{2w+1} \sum_{k=-w}^{w} |x_{i+k} - \bar{x}_i|, \quad \bar{x}_i = \frac{1}{2w+1} \sum_{k=-w}^{w} x_{i+k} \tag{5}$$

where $\bar{x}_i$ is the local mean. Segments where $\sigma_i$ exceeds an adaptive threshold $T_\sigma$ (defined analogously to $T_E$) are flagged as candidate intervals. As demonstrated in Figure3, this dispersion feature is highly sensitive to closely spaced keystrokes, efectively mitigating false negatives ofthe energy-based method.

Candidate segments derived from energy and dispersion metrics remain susceptible to environmental noise. To ensure robust segmentation, we propose a cross-modal alignment strategy using acoustic anchors, precise keystroke timestamps extracted from audio (Sec. 4.4). By synchronizing both modalities, we evaluate the combined candidate pool and map each acoustic anchor to the CSI segment whose midpoint is temporally closest. This matched segment becomes the definitive spatial representation for the keystroke, while surplus unanchored segments are definitively discarded. This multimodal fusion rigorously filters out artifacts, ensuring highly accurate segmentation.

## 4.4 Spatial–Temporal Feature Analysis

With individual keystroke CSI segments isolated, we extract their spatial signatures. A distinct physical advantage of AIRKEy is that our ACK-induction mechanism transforms the victim's own device into the sensing transmitter. Consequently, finger micro-motions occur in the immediate near-field of the antennas, yielding significantly higher-quality spatial perturbations compared to externally deployed transceivers [59]. Because diferent keys dictate distinct finger trajectories, whereas repeated presses of the same key produce highly correlated physical perturbations, we can cluster keystrokes based on CSI waveform similarity.

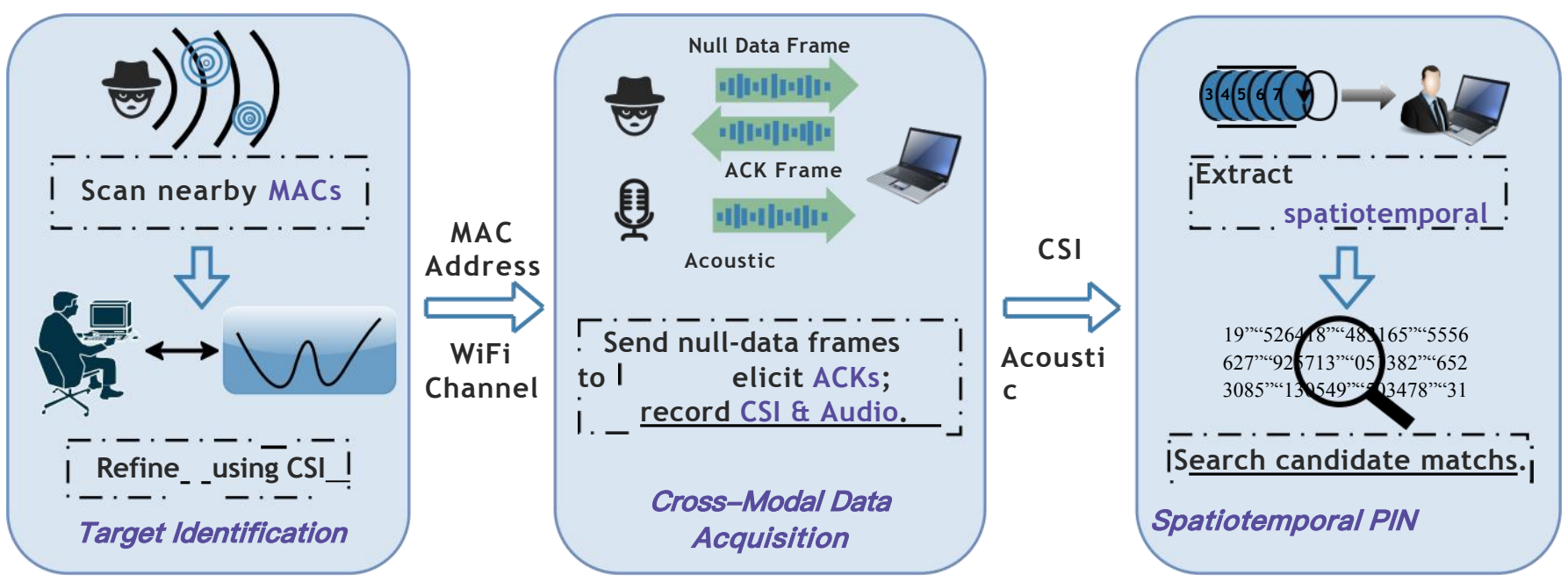

**Figure 2: The attack processs of AIRKEy.**

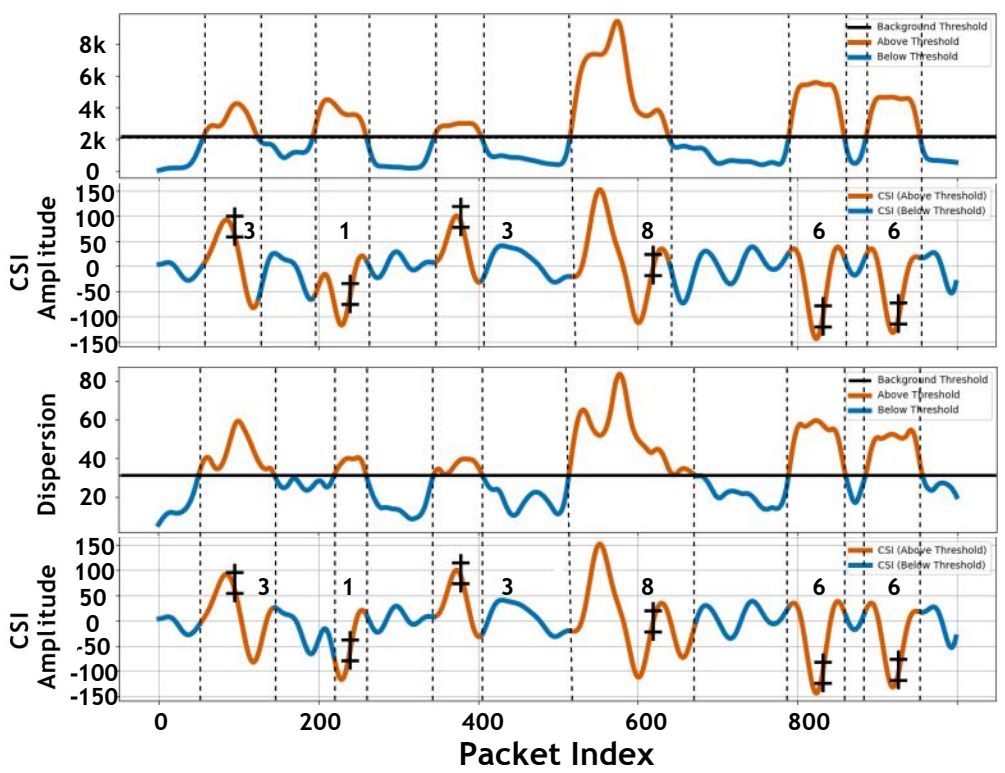

**Figure 3: CSI waveform segmentation**

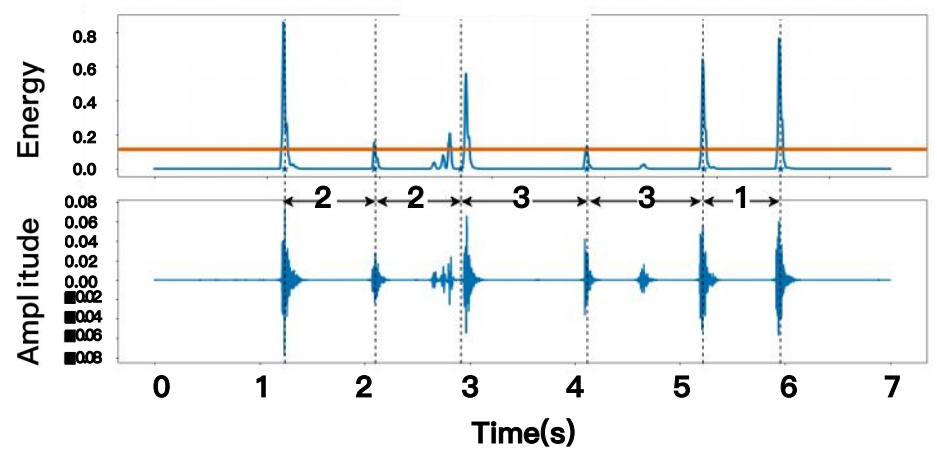

**Figure 4: Energy curve of the acoustic signal.**

To quantify this spatial consistency, we apply Dynamic Time Warping (DTW) to compute the pairwise similarity, $D_s(A, B) = DTW(A, B)$, efectively accommodating slight temporal variations in typing speed. Using an empirical threshold ($T_s = 3$), we employ a greedy clustering approach to assign identical labels to highly similar waveform segments, thereby mapping the input sequence to an abstract spatial template. For instance, as illustrated in Figure 3, the PIN sequence "313866" exhibits low DTW costs (high similarity) between the 1st/3rd segments (cost: 2.24) and the 5th/6th segments (cost: 1.46), with all other pairwise costs exceeding 10. Thus, AIRKEy maps "313866" to the abstract spatial feature "121344". This spatial abstraction drastically constrains the guessing space. For a 6-digit PIN, it maps the $10^6$ raw permutations into merely 203 unique spatial templates. By accurately recovering this spatial pattern, the candidate pool of numeric sequences is exponentially reduced from one million to an average of just 4,926 combinations.

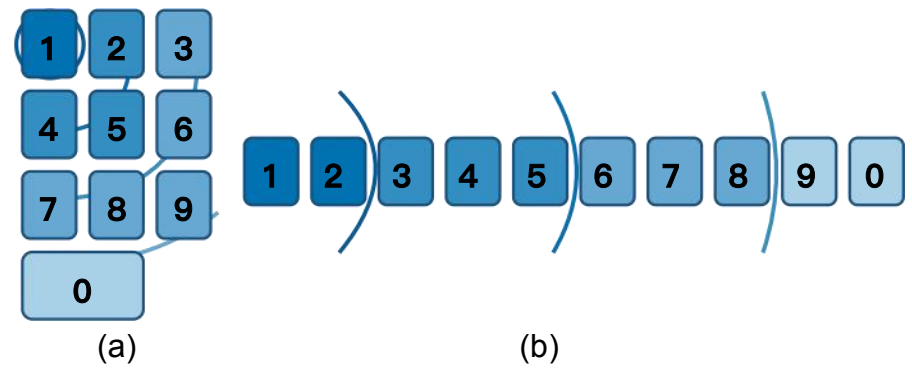

**Figure 5: Diferent keyboard layouts**

To extract precise temporal features, we process the microphone recordings using time-diference delay-and-sum beamforming [4], which suppresses of-axis noise and isolates the target typing sounds. Following standard spectral denoising, we compute the short-time energy $e_i$ using a sliding window:

$$e_i = \frac{1}{2w + 1} \sum_{k=-w}^{w} |s_{i+k}|^2 \quad (6)$$

where $s$ denotes the audio amplitude. We first identify candidate peaks whose energy values exceed an empirical threshold defined as the 95th percentile of the energy distribution. Subsequently, K-means clustering is applied to these prominent peaks (Fig. 4) to reliably determine the exact timestamp $t_m$ for each keystroke.

Assuming muscle-memory typing, the time interval between consecutive keystrokes is physically correlated with the spatial distance between the keys. We define the normalized temporal distance between two inter-keystroke intervals, $I_A$ and $I_B$, as:

$$D_t(I_A, I_B) = \frac{|I_A - I_B|}{\max(I_A, I_B)} \quad (7)$$

Interval pairs with $D_t < 0.15$ are clustered into the same duration class. We then rank these classes by their mean lengths to generate an abstract temporal feature vector (e.g., assigning ascending labels 1, 2, 3).

As illustrated in Figure 5, diferent keyboard layouts yield distinct temporal signatures. Focusing on the ubiquitous linear number row (Fig. 5(b)), we explicitly map these ranked duration classes to physical key separations (e.g., adjacent/same keys, 1–3, 4–6, and 7–8 positions apart). Consequently, a PIN sequence like "313921" systematically translates into the temporal feature "22331". Acoustic temporal features alone efectively capture typing rhythms, condensing the $10^6$ raw permutations into 415 distinct patterns (averaging 2,410 candidates). However, the true potency of AIRKEy lies in its cross-modal alignment. By strictly intersecting the CSI-derived spatial templates with the acoustic-guided temporal constraints, we jointly filter out unimodal ambiguities. Under ideal conditions, this spatiotemporal fusion collapses the final guessing space from one million to merely a few dozen combinations, enabling highly efficient, zero-training PIN inference.

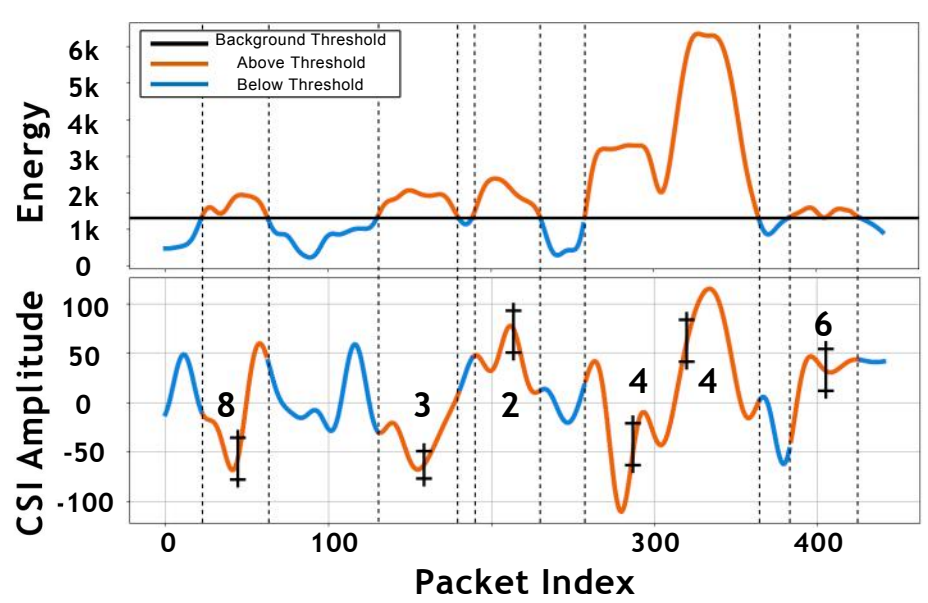

Figure 6: Keystroke waveform fusion

## 4.5 Cross-Modal Disentanglement of Waveform Fusion

When a user types rapidly, multipath propagation delays can cause the CSI attenuation of consecutive keystrokes to superpose, creating a phenomenon we term *waveform fusion* (Figure6). Unimodal WiFi methods typically misclassify these merged waveforms as a single keystroke, causing fatal sequence misalignment and rendering password recovery impossible [20, 59]. To address this physical bottleneck, AIRKEy leverages cross-modal mismatch detection. Specifically, if two distinct acoustic timestamps fall within a single continuous CSI segment, a fusion event is explicitly flagged. Rather than discarding this merged segment or allowing it to corrupt the spatial similarity computation, we isolate it and transform the fusion artifact into a definitive spatiotemporal constraint. Temporally, the fused interval is used to recalibrate the minimum interval threshold for the entire sequence, correcting potential temporal classification errors. Spatially, the fused segment is masked (marked as unavailable for standard DTW), but constrained by a strict neighborhood rule: the two fused keystrokes must represent either the exact same key or two physically adjacent keys.

By incorporating these fusion-specific constraints, AIRKEy substantially preserves inference robustness. Consider the PIN sequence "832446", which ideally yields spatial and temporal features resolving to just 498 candidates. If the 2nd and 3rd keystrokes fuse, existing WiFi-only methods extract corrupted, shortened features (e.g., spatial "12345"), leading to total inference failure. Even if a baseline system knows a fusion occurred but lacks our cross-modal neighborhood constraints, the resulting ambiguity expands the search space to 3,194 candidates. By contrast, AIRKEy explicitly encodes the fused pair with a localized constraint (represented abstractly as spatial "123X4", where 'X' enforces adjacency), effectively bounding the candidate set to only 1,680. Ultimately, while pure RF sensing degrades under severe waveform fusion and pure acoustic sensing demands intrusive microphone arrays for spatial localization [21], AIRKEy's multimodal architecture elegantly sidesteps both limitations. By strictly utilizing audio for temporal alignment and CSI for spatial signatures, we transform previously catastrophic waveform fusions into exploitable inference constraints, achieving robust keystroke recognition with minimal hardware complexity.

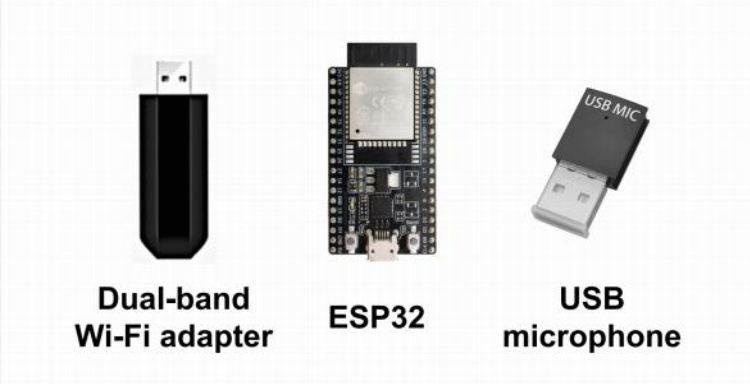

Figure 7: Equipments used in AIRKEy.

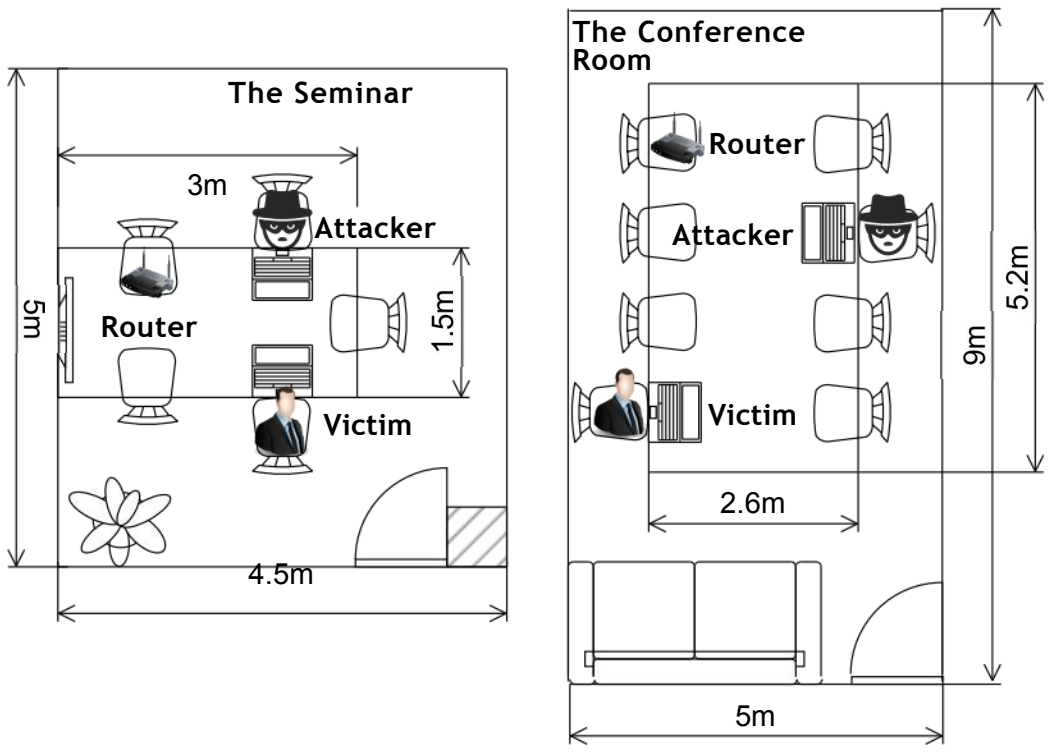

Figure 8: Evaluation environments.

# 5 Implementation and Evaluation

## 5.1 Prototype and Evaluation Setup

As shown in Figure 7, the AIRKEy prototype is constructed entirely from unobtrusive, commercial of-the-shelf (COTS) components connected via USB to the attacker's laptop. A standard WiFi adapter handles target discovery and frame injection; a low-cost ESP32 microcontroller passively extracts CSI from the induced ACKs; and two lightweight microphones capture the acoustic temporal anchors. This compact configuration ensures high portability and stealth without requiring specialized transceivers.

Table 1: Devices used in the evaluation.

| Device Name | ACK-induced rate |
|---|---|
| Lenovo Legion Y9000P | 104.76 packets/s |
| Lenovo Legion R7000 | 102.05 packets/s |
| Lenovo Legion Y7000 | 102.07 packets/s |
| ASUS TUF Gaming A15 (2023) | 103.69 packets/s |
| HUAWEI MateBook 13 | 106.98 packets/s |
| HUAWEI MateBook 14 | 104.91 packets/s |
| MECHREVO Unbounded 15 Pro | 103.55 packets/s |
| MECHREVO Wing Dragon 15 Pro | 102.90 packets/s |
| MECHREVO Aurora Pro | 104.38 packets/s |

We comprehensively evaluated AIRKEy in two real-world environments (Figure 8): a quiet seminar room and a dynamic conference room (featuring bystander motion, multipath interference, and 40–50 dB ambient noise). All CSI data were collected under non-line-of-sight (NLOS) conditions, where the attacker could not directly observe the keyboard being typed by the volunteers. Because most modern laptops lack a dedicated numpad, our primary evaluation focuses on the ubiquitous linear number row (3×3 keypad evaluations are deferred to Sec. 5.3). To ensure ethical compliance, 10 volunteers participated by typing 6-digit PINs [20, 59] randomly assigned from a pool of 177 sequences. To guarantee ecological validity, participants extensively practiced their assigned PINs to emulate rapid, muscle-memory typing. In total, we collected 720 continuous traces across 9 diverse commodity laptops (Table 1). Crucially, under a transmission rate of 200 induced null data frames per second, empirical measurements confirmed that the induced-ACK rate across these unmodified targets consistently stabilized at roughly 100-110 packets/s.

## 5.2 Overall Performance

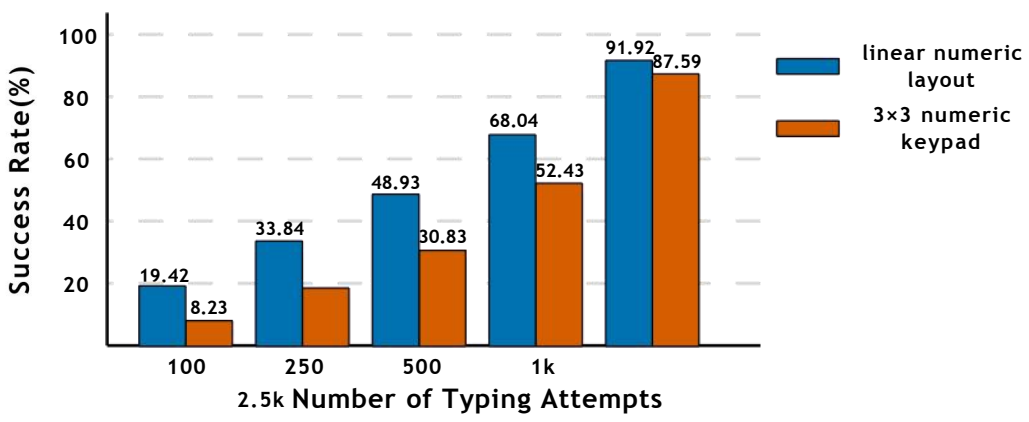

**Figure 9: Performance under diferent attack intensities and layouts.**

Figure9evaluates the overall top-$K$ accuracy of AIRKEy, defined as the fraction of trials where the ground-truth PIN ranks within the top $K$ candidates. From a single observation, AIRKEy achieves 19.42% accuracy at $K = 100$, steadily scaling to 48.93% ($K = 500$), 68.04% ($K = 1,000$), and 91.92% ($K = 2,500$). Because most authen - tication interfaces enforce strict account lockouts (e.g., $< 10$ failed attempts), this single-trace performance poses a constrained imme- diate threat. While these success rates appear lower than existing baselines [9, 59], this discrepancy stems directly from our highly stringent evaluation protocol. Prior evaluations [20, 59] typically featured slow, deliberate typing by unpracticed users. In contrast, our stringent protocol mandated rapid, muscle-memory PIN entry.

## 5.3 Impact of Practical Factors

**Impact of Volunteers.** As depicted in Figure10, inference accuracy naturally fluctuates across participants due to typing habits. Observational analysis reveals that users exhibiting erratic key-press forces or inconsistent rhythmic intervals introduce unintended distortions into both modalities, which slightly degrades the discriminative power of the resulting spatiotemporal features.

**Impact of Environments.** AIRKEy demonstrates strong envi - ronmental resilience. In the quiet, short-range seminar room, the top-2,500 accuracy reaches 92.98%. Impressively, in the highly chal- lenging conference room (longer distance, severe multipath, and 40–50 dB ambient noise), accuracy experiences only a marginal drop to 90.88%. While intense ambient noise slightly perturbs the acoustic temporal anchors and dynamic multipath weakens the CSI spatial signatures, our cross-modal fusion efectively absorbs most of these environmental degradations.

**Impact of CSI Waveform Fusion.** To quantify the necessity of our cross-modal disentanglement mechanism, Figure 11compares AIRKEy’s performance on waveform-fusion samples under two settings: with fusion handling (left bars) and with fusion detection only (right bars). In the latter case, failure to correctly identify fusion events directly leads to inference failure due to the absence of efective constraints. During rapid PIN entry, ignoring fused

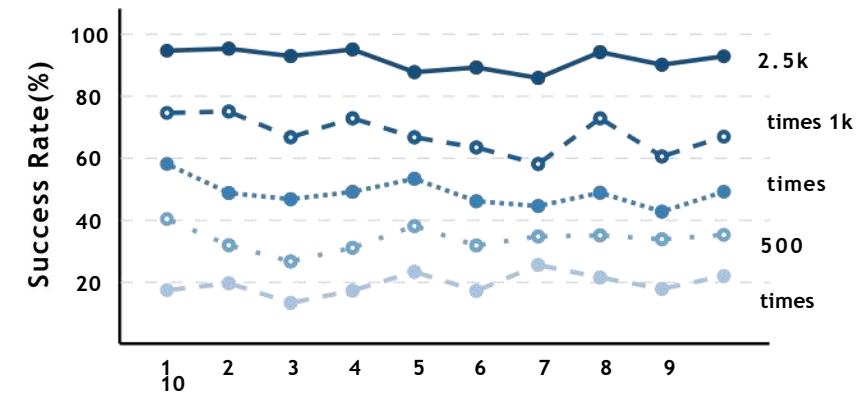

**Figure 10: Comparison of results among diferent volunteers.**

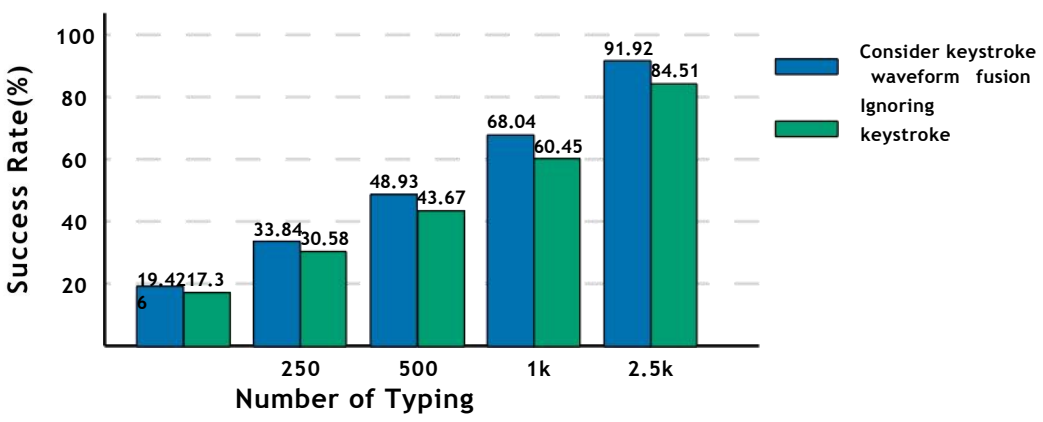

**Figure 11: Performance with and without considering wave – form fusing.**

**Table 2: Comparison experiments.**

| Method | 100 guesses | 250 guesses | 500 guesses | 1000 guesses | 2500 guesses |
|---|---|---|---|---|---|
| WINK | 0.0420 | 0.0918 | 0.1490 | 0.2320 | 0.4049 |
| WiPass (in-domain) | 0.2332 | 0.3687 | 0.5550 | 0.7364 | 0.9694 |
| WiPass (out-of-domain) | 0.0020 | 0.0049 | 0.0099 | 0.0197 | 0.0494 |
| KAN-Sense (in-domain) | 0.1965 | 0.3842 | 0.5635 | 0.7036 | 0.9275 |
| KAN-Sense (out-of-domain) | 0.0014 | 0.0035 | 0.0071 | 0.0141 | 0.0353 |
| AIRKEy | 0.1942 | 0.3384 | 0.4893 | 0.6804 | 0.9192 |

CSI waveforms causes severe spatial misalignment. In contrast, by leveraging acoustic anchors to explicitly detect fused segments and imposing spatiotemporal constraints, AIRKEy substantially recovers inference accuracy, validating the robustness of our multimodal architecture.

**Impact of 3×3 Layout.** We benchmarked the standard linear number row against a denser 3×3 numeric keypad (Figure9). The 3×3 layout yields an average accuracy drop of 12.99%. This degradation is fundamentally geometrical: the compact grid drastically reduces the physical distance between distinct keys, which inherently diminishes the distinctiveness of both the CSI spatial perturbations and the acoustic temporal intervals.

## 5.4 Comparative Analysis and Ablation

For a rigorous comparison, we evaluate AIRKEy against WINK [59], a representative zero-training CSI inference framework. We enhance WINK with our ACK-induced CSI acquisition and preprocessing pipeline, yielding a strong CSI-only baseline that also serves as a modality ablation. As shown in Table2, AIRKEy achieves 33.84% accuracy within 250 guesses and 91.92% within 2,500 guesses, compared with 9.18% and 40.49% for WINK, respectively. Despite using the same high-quality CSI, WINK remains vulnerable to segmentation errors during rapid typing, whereas the acoustic anchors in AIRKEy separate temporally fused keystrokes. The theoretical lower bound derived from our spatiotemporal abstraction averages 370 guesses; in practice, AIRKEy requires 1,742 guesses on average, substantially fewer than WINK’s 12,060. These results demonstrate that cross-modal fusion is essential for overcoming the limitations of CSI-only inference.

We further compare AIRKEy with two state-of-the-art learning-based methods, WiPass [44] and KAN-Sense [27]. With sufficient user- and device-specific training data, they achieve 96.94% and 92.75% accuracy within 2,500 guesses, respectively, slightly exceeding AIRKEy's 91.92%. However, their accuracy drops to 4.94% and 3.53% for unseen users and devices, revealing their dependence on scenario-specific training data. In contrast, AIRKEy requires no prior data from target users, devices, or environments. Its combination of CSI-based spatial signatures and acoustic temporal anchors provides stable zero-training inference across scenarios. Although slightly less accurate than fully trained models in-domain, AIRKEy avoids costly data collection and retraining while ofering substantially stronger cross-domain robustness and practical applicability.

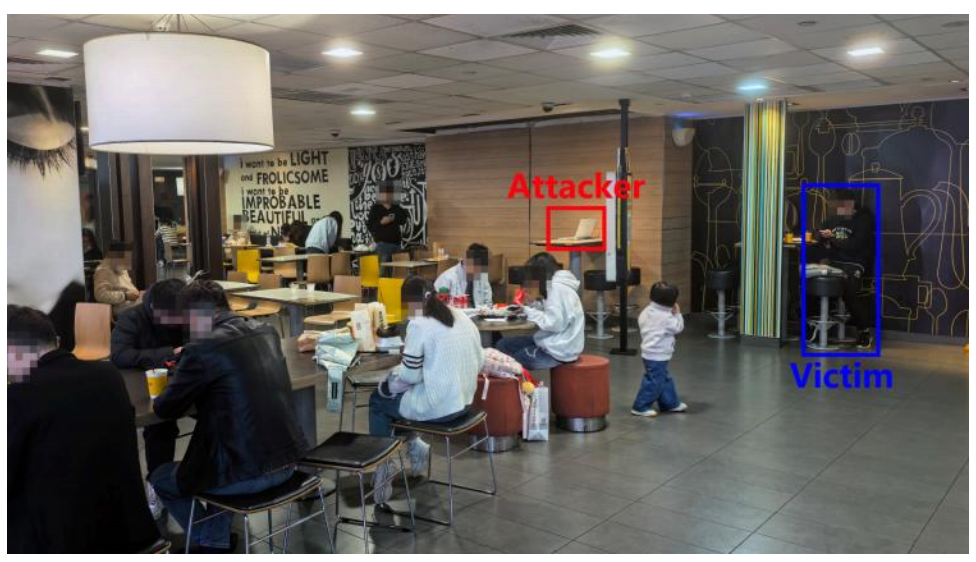

Figure 12: Evaluation environment of the McDonald.

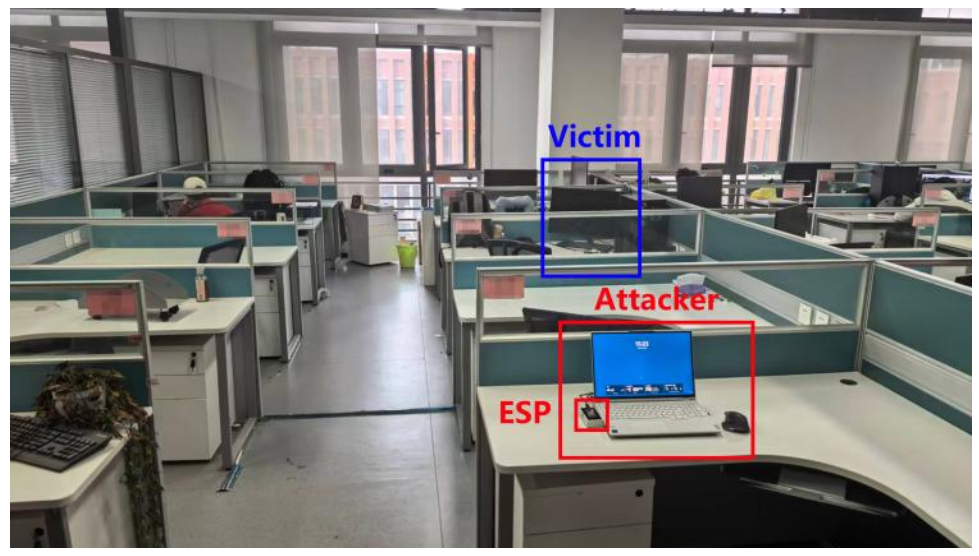

Figure 13: Evaluation environment of the office.

## 5.5 Real-World Case Studies

While single-trace inference under strict real-world conditions poses a constrained immediate threat, practical adversaries can easily exploit routine user habits to aggregate multiple observation windows. To evaluate the efficacy of this multi-trace fusion strategy, we conducted two case studies in practical environments.

**Case Study 1: Noisy Public Environment (McDonald's).** We deployed the attack in a crowded McDonald's during peak lunch hours (Figure 12). The attacker concealed the compact hardware setup and isolated the victim's MAC address using the previously described physical-activity correlation. The attacker then monitored the victim's routine laptop unlocks (e.g., left their seat multiple times to order food, pick it up, and use the restroom, then returning to their seat), capturing three independent AIRKEy traces. As summarized in Table 3, environmental dynamics introduced anomalies in Trace 3 (spatial) and Trace 2 (temporal). However, by jointly leveraging the spatial signature [1, 2, 3, 4, 2, 5] and two candidate temporal signatures [2, 1, 3, 3, 4] and [1, 2, 3, 3, 4], the attacker efectively mitigates uncertainty. This cross-modal constraint reduces the search space to 56 candidates, recovering the ground-truth PIN "312719" on the 6th guess (testing sequence: 097291, 124920, 213810, 213819, 312710, **312719**).

Table 3: Spatial and temporal features corresponding to three password-entry session.

| Keystroke Behavior | Spatial Feature | Temporal Feature |
|---|---|---|
| 1 | [1,2,3,4,2,5] | [1,2,3,3,4] |
| 2 | [1,2,3,4,2,5] | [2,1,3,4,4] |
| 3 | [1,2,3,4,5,6] | [2,1,3,3,4] |

**Case Study 2: Shared Office Environment.** The second study targeted a colleague in an office setting (Figure 13). Over a half-day period, the attacker discreetly captured five unlocking traces as the victim periodically left and returned to their desk. The extracted features (Table4) reveal a critical scenario: Trace 4 exhibited a waveform fusion event. Rather than causing inference failure, AIRKEy robustly handled this by encoding the fusion as an explicit adjacency constraint ('X'). Through cross-trace aggregation and majority voting, the attacker efectively mitigated trace-specific errors and extracted the consensus spatial signature [1, 2, 3, 4, 2, 5] and temporal signature [2, 1, 3, 3, 4]. The system further obtained consistent spatiotemporal features with those derived in Case 1. Consequently, the search space was further constrained, enabling the attacker to recover the PIN "312719" on merely the 2nd attempt (testing sequence: 312710, **312719**, 312810).

Table 4: Spatial and temporal features corresponding to five password-entry session.

| Keystroke Behavior | Spatial Feature | Temporal Feature |
|---|---|---|
| 1 | [1,2,3,4,2,5] | [1,2,3,3,4] |
| 2 | [1,2,3,4,2,5] | [2,1,3,3,4] |
| 3 | [1,2,3,4,5,6] | [2,1,3,3,4] |
| 4 | [1,X,2,3,4] | [2,1,3,3,4] |
| 5 | [1,2,3,4,2,5] | [2,1,3,3,4] |

Ultimately, these evaluations highlight the compounding danger of multimodal eavesdropping. By aggregating cross-modal spatiotemporal features across multiple natural observations, an adversary efectively filters out unimodal anomalies and exponentially collapses the guessing space. This multi-trace fusion unequivocally escalates AIRKEy from a constrained single-attempt vulnerability into a highly practical and severe security threat.

# 6 Conclusion

This paper introduces AIRKEy, a stealthy, cross-modal eavesdropping framework that fundamentally broadens the threat landscape of passive WiFi sensing. We demonstrate that non-line-of-sight attackers can covertly induce ACK responses from unassociated, unmodified target devices to extract CSI signals. Crucially, by integrating an auxiliary acoustic modality to provide temporal anchors, AIRKEy effectively disentangles waveform fusion, enabling highly robust, zero-training keystroke inference. Extensive real-world case studies in uncontrolled environments reveal that through multi-trace spatiotemporal aggregation, AIRKEy can successfully recover 6-digit PINs in as few as 6 guesses. Ultimately, this work exposes a severe, practical vulnerability in ubiquitous smart interfaces: the unauthenticated leakage of ACK-induced CSI, when exploited via multimodal fusion, constitutes a potent attack vector that demands immediate protocol- and hardware-level mitigations.

## Acknowledgments

This work is supported by the National Natural Science Foundation of China (No. 62502482, No. 62372149, No. U23A20303, and No. 62572168). It is also supported by the Anhui Provincial Natural Science Foundation 2508085MF151.

## References

[1] Ali Abedi, Haofan Lu, Alex Chen, Charlie Liu, and Omid Abari. 2023. WiFi physical layer stays awake and responds when it should not. *IEEE Internet of Things Journal* 11, 3 (2023), 4483–4496.
[2] Kamran Ali, Alex X Liu, Wei Wang, and Muhammad Shahzad. 2015. Keystroke recognition using wifi signals. In *Proceedings of the 21st annual international conference on mobile computing and networking*. 90–102.
[3] Kiran Balagani, Matteo Cardaioli, Mauro Conti, Paolo Gasti, Martin Georgiev, Tristan Gurtler, Daniele Lain, Charissa Miller, Kendall Molas, Nikita Samarin, et al. 2019. Pilot: Password and pin information leakage from obfuscated typing videos. *Journal of Computer Security* 27, 4 (2019), 405–425.
[4] Jacob Benesty, Jingdong Chen, and Yiteng Huang. 2008. *Microphone Array Signal Processing*. Springer Topics in Signal Processing, Vol. 1. Springer Berlin Heidelberg, Berlin, Heidelberg. doi:10.1007/978-3-540-78612-2_9
[5] P. Biondi. 2020. Scapy. https://scapy.net/.
[6] Liang Cai and Hao Chen. 2011. TouchLogger: Inferring Keystrokes on Touch Screen from Smartphone Motion. In *6th USENIX Workshop on Hot Topics in Security (HotSec 11)*. USENIX Association, San Francisco, CA.
[7] Matteo Cardaioli, Stefano Cecconello, Mauro Conti, Simone Milani, Stjepan Picek, and Eugen Saraci. 2022. Hand me your PIN! inferring ATM PINs of users typing with a covered hand. In *31st USENIX Security Symposium (USENIX Security 22)*. 1687–1704.
[8] Hao Chen, Penghao Wang, Feng Hong, Zhongwen Guo, and Chao Liu. 2025. BEyes: Unseen Eyes Snooping Pattern Lock via BFI. In *2025 IEEE 45th International Conference on Distributed Computing Systems (ICDCS)*. IEEE, 122–132.
[9] Siyu Chen, Hongbo Jiang, Jingyang Hu, Tianyue Zheng, Mengyuan Wang, Zhu Xiao, Daibo Liu, and Jun Luo. 2024. Echoes of fingertip: unveiling POS terminal passwords through Wi-Fi beamforming feedback. *IEEE Transactions on Mobile Computing* (2024).
[10] Yimin Chen, Tao Li, Rui Zhang, Yanchao Zhang, and Terri Hedgpeth. 2018. Eyetell: Video-assisted touchscreen keystroke inference from eye movements. In *2018 IEEE Symposium on Security and Privacy (SP)*. IEEE, 144–160.
[11] Patrick Cronin, Xing Gao, Chengmo Yang, and Haining Wang. 2021. Charger-Surfing: Exploiting a Power Line Side-Channel for Smartphone Information Leakage. In *30th USENIX Security Symposium (USENIX Security 21)*. USENIX Association, 681–698.
[12] Song Fang, Ian Markwood, Yao Liu, Shangqing Zhao, Zhuo Lu, and Haojin Zhu. 2018. No training hurdles: Fast training-agnostic attacks to infer your typing. In *Proceedings of the 2018 ACM SIGSAC Conference on Computer and Communications Security*. 1747–1760.
[13] Yunpeng Feng, Daibo Liu, Wenqiang Jin, and Liangyi Gong. 2025. KeyPrint: Practical Black-box Keystroke Inference Attacks to Mobile Devices. *Proceedings of the ACM on Interactive, Mobile, Wearable and Ubiquitous Technologies* 9, 2 (2025), 1–30.
[14] Francesco Gringoli, Matthias Schulz, Jakob Link, and Matthias Hollick. 2019. Free your CSI: A channel state information extraction platform for modern Wi-Fi chipsets. In *Proceedings of the 13th International Workshop on Wireless Network Testbeds, Experimental Evaluation & Characterization*. 21–28.
[15] Yu Gu, Xiang Zhang, Yantong Wang, Meng Wang, Huan Yan, Yusheng Ji, Zhi Liu, Jianhua Li, and Mianxiong Dong. 2022. WiGRUNT: WiFi-enabled gesture recognition using dual-attention network. *IEEE Transactions on Human–Machine Systems* 52, 4 (2022), 736–746.
[16] Yu Gu, Xiang Zhang, Huan Yan, Jingyang Huang, Zhi Liu, Mianxiong Dong, and Fuji Ren. 2023. WiFE: WiFi and vision based unobtrusive emotion recognition via gesture and facial expression. *IEEE Transactions on Affective Computing* 14, 4 (2023), 2567–2581.
[17] Daniel Halperin, Wenjun Hu, Anmol Sheth, and David Wetherall. 2011. Tool release: Gathering 802.11 n traces with channel state information. *ACM SIGCOMM computer communication review* 41, 1 (2011), 53–53.
[18] Steven M. Hernandez and Eyuphan Bulut. 2020. Lightweight and Standalone IoT Based WiFi Sensing for Active Repositioning and Mobility. In *21st International Symposium on "A World of Wireless, Mobile and Multimedia Networks" (WoWMoM) (WoWMoM 2020)*. Cork, Ireland.
[19] Jingyang Hu, Hongbo Jiang, Tianyue Zheng, Jingzhi Hu, Hongbo Wang, Hangcheng Cao, Zhe Chen, and Jun Luo. 2024. M2-Fi: Multi-person Respiration Monitoring via Handheld WiFi Devices. In *Proceedings of the IEEE INFOCOM 2024*. IEEE, 1221–1230.
[20] Jingyang Hu, Hongbo Wang, Tianyue Zheng, Jingzhi Hu, Zhe Chen, Hongbo Jiang, and Jun Luo. 2023. Password-stealing without hacking: Wi-Fi enabled practical keystroke eavesdropping. In *Proceedings of the 2023 ACM SIGSAC conference on computer and communications security*. 239–252.
[21] Jinyang Huang, Jia-Xuan Bai, Xiang Zhang, Zhi Liu, Yuanhao Feng, Jianchun Liu, Xiao Sun, Mianxiong Dong, and Meng Li. 2024. Keystrokesnifer: An of-the-shelf smartphone can eavesdrop on your privacy from anywhere. *IEEE Transactions on Information Forensics and Security* 19 (2024), 6840–6855.
[22] IEEE. 2021. IEEE Standard for Information Technology–Telecommunications and Information Exchange between Systems - Local and Metropolitan Area Networks–Specific Requirements - Part 11: Wireless LAN Medium Access Control (MAC) and Physical Layer (PHY) Specifications. *IEEE Std 802.11–2020 (Revision of IEEE Std 802.11–2016)* (2021), 1–4379. doi:10.1109/IEEESTD.2021.9363693
[23] Fidelity Investments Inc. 2021. Fidelity's Automated Service Telephone. https://www.fidelity.com/customer-service/phone-numbers/fast/overview.
[24] Songming Jia, Yan Lu, Bin Liu, Xiang Zhang, Peng Zhao, Xinmeng Tang, Yelin Wei, Jinyang Huang, Huan Yan, and Zhi Liu. 2026. Breaking Coordinate Overfitting: Geometry-Aware WiFi Sensing for Cross-Layout 3D Pose Estimation. *arXiv preprint arXiv:2601.12252* (2026).
[25] Wenqiang Jin, Srinivasan Murali, Huadi Zhu, and Ming Li. 2021. Periscope: A keystroke inference attack using human coupled electromagnetic emanations. In *Proceedings of the 2021 ACM SIGSAC Conference on Computer and Communications Security*. 700–714.
[26] Tejas Kannan, Synthia Wang, Max Sunog, Abe Bueno de Mesquita, Nick Feamster, and Henry Hofmann. 2024. Acoustic keystroke leakage on smart televisions. In *Network and Distributed System Security Symposium. Internet Society*.
[27] Min Koo and Jonghyun Park. 2025. KAN-Sense: Keypad Input Recognition via CSI Feature Clustering and KAN-Based Classifier. *Electronics* 14, 15 (2025), 2965. doi:10.3390/electronics14152965
[28] Jiachun Li, Yan Meng, Fazhong Liu, Tian Dong, Suguo Du, Guoxing Chen, Yuling Chen, and Haojin Zhu. 2025. Synergistic Multi-Modal Keystroke Eavesdropping in Virtual Reality With Vision and Wi-Fi. *IEEE Transactions on Information Forensics and Security* (2025).
[29] Mengyuan Li, Yan Meng, Junyi Liu, Haojin Zhu, Xiaohui Liang, Yao Liu, and Na Ruan. 2016. When CSI meets public WiFi: Inferring your mobile phone password via WiFi signals. In *Proceedings of the 2016 ACM SIGSAC conference on computer and communications security*. 1068–1079.
[30] Xin Li, Jingzhi Hu, Hongbo Wang, Zhe Chen, and Jun Luo. 2025. Enabling Ultra-Wideband Wi-Fi Sensing via Sparse Channel Sampling. *IEEE Journal on Selected Areas in Communications* (2025).
[31] Jinyi Liu, Wenwei Li, Tao Gu, Ruiyang Gao, Bin Chen, Fusang Zhang, Dan Wu, and Daqing Zhang. 2023. Towards a dynamic fresnel zone model to wifi-based human activity recognition. *Proceedings of the ACM on Interactive, Mobile, Wearable and Ubiquitous Technologies* 7, 2 (2023), 1–24.
[32] Xiangyu Liu, Zhe Zhou, Wenrui Diao, Zhou Li, and Kehuan Zhang. 2015. When good becomes evil: Keystroke inference with smartwatch. In *Proceedings of the 22nd ACM SIGSAC Conference on Computer and Communications Security*. 1273–1285.
[33] Li Lu, Jiadi Yu, Yingying Chen, Yanmin Zhu, Xiangyu Xu, Guangtao Xue, and Minglu Li. 2019. Keylistener: Inferring keystrokes on qwerty keyboard of touch screen through acoustic signals. In *IEEE INFOCOM 2019–IEEE Conference on Computer Communications*. IEEE, 775–783.
[34] Federico Maggi, Alberto Volpatto, Simone Gasparini, Giacomo Boracchi, and Stefano Zanero. 2011. A fast eavesdropping attack against touchscreens. In *2011 7th International Conference on Information Assurance and Security (IAS)*. IEEE, 320–325.
[35] Anindya Maiti, Oscar Armbruster, Murtuza Jadliwala, and Jibo He. 2016. Smartwatch-based keystroke inference attacks and context-aware protection mechanisms. In *Proceedings of the 11th ACM on Asia Conference on Computer and Communications Security*. 795–806.
[36] Philip Marquardt, Arunabh Verma, Henry Carter, and Patrick Traynor. 2011. (sp) iphone: Decoding vibrations from nearby keyboards using mobile phone accelerometers. In *Proceedings of the 18th ACM conference on Computer and communications security*. 551–562.
[37] Xuanqi Meng, Jiarun Zhou, Xiulong Liu, Xinyu Tong, Wenyu Qu, and Jianrong Wang. 2023. Secur-Fi: A secure wireless sensing system based on commercial Wi-Fi devices. In *IEEE INFOCOM 2023–IEEE Conference on Computer Communications*. IEEE, 1–10.
[38] Yan Meng, Jinlei Li, Haojin Zhu, Xiaohui Liang, Yao Liu, and Na Ruan. 2020. Revealing your mobile password via WiFi signals: Attacks and countermeasures. *IEEE Transactions on Mobile Computing* 19, 2 (2020), 432–449.
[39] Emiliano Miluzzo, Alexander Varshavsky, Suhrid Balakrishnan, and Romit Roy Choudhury. 2012. Tapprints: your finger taps have fingerprints. In *Proceedings of the 10th international conference on Mobile systems, applications, and services*. 323–336.
[40] John V Monaco. 2018. Sok: Keylogging side channels. In *2018 IEEE Symposium on Security and Privacy (SP)*. IEEE, 211–228.
[41] Min Peng, Xianxin Fu, Haiyang Zhao, Yu Wang, and Caihong Kai. 2024. LiKey: Location-independent keystroke recognition on numeric keypads using WiFi signal. *Computer Networks* 245 (2024), 110354.

[42] Rahul Raguram, Andrew M White, Dibyendusekhar Goswami, Fabian Monrose, and Jan-Michael Frahm. 2011. iSpy: automatic reconstruction of typed input from compromising reflections. In *Proceedings of the 18th ACM conference on Computer and communications security*. 527–536.

[43] Mohd Sabra, Anindya Maiti, and Murtuza Jadliwala. 2020. Zoom on the keystrokes: Exploiting video calls for keystroke inference attacks. *In Network and Distributed Systems Security (NDSS) Symposium*. (2020).

[44] Xuan Shen, Zhen Ni, Lei Liu, Jian Yang, and Kamal Ahmed. 2021. WiPass: 1D-CNN-based smartphone keystroke recognition using WiFi signals. *Pervasive and Mobile Computing* 73 (2021), 101393. doi:10.1016/j.pmcj.2021.101393

[45] David Slater, Scott Novotney, Jessica Moore, Sean Morgan, and Scott Tenaglia. 2019. Robust keystroke transcription from the acoustic side-channel. In *Proceedings of the 35th Annual Computer Security Applications Conference*. 776–787.

[46] Jingchao Sun, Xiaocong Jin, Yimin Chen, Jinxue Zhang, Yanchao Zhang, and Rui Zhang. 2016. Visible: Video-assisted keystroke inference from tablet backside motion.. In *NDSS*.

[47] Xinyu Tong, Weiping Ge, Yichen Tian, Zijuan Liu, Xiulong Liu, and Wenyu Qu. 2024. NNE-tracking: A neural network enhanced framework for device-free Wi-Fi tracking. *IEEE Transactions on Mobile Computing* 23, 9 (2024), 8981–8998.

[48] Yazhou Tu, Liqun Shan, Md Imran Hossen, Sara Rampazzi, Kevin Butler, and Xiali Hei. 2023. Auditory Eyesight: Demystifying $\mu$s-Precision Keystroke Tracking Attacks on Unconstrained Keyboard Inputs. In *32nd USENIX Security Symposium (USENIX Security 23)*. USENIX Association, 175–192.

[49] Martin Vuagnoux and Sylvain Pasini. 2009. Compromising electromagnetic emanations of wired and wireless keyboards.. In *USENIX security symposium*, Vol. 8. 1–16.

[50] Chen Wang, Xiaonan Guo, Yan Wang, Yingying Chen, and Bo Liu. 2016. Friend or foe? Your wearable devices reveal your personal pin. In *Proceedings of the 11th ACM on Asia conference on computer and communications security*. 189–200.

[51] Hongbo Wang, Jingyang Hu, Tianyue Zheng, Jingzhi Hu, Zhe Chen, Hongbo Jiang, Yuanjin Zheng, and Jun Luo. 2024. MuKI-Fi: Multi-person keystroke inference with BFI-enabled Wi-Fi sensing. *IEEE Transactions on Mobile Computing* 23, 10 (2024), 9835–9850.

[52] Penghao Wang, Jingzhi Hu, Chao Liu, and Jun Luo. 2024. RefleXnoop: Passwords Snooping on NLoS Laptops Leveraging Screen-Induced Sound Reflection. In *Proceedings of the 2024 on ACM SIGSAC Conference on Computer and Communications Security*. 3361–3375.

[53] Yao Wang, Wandong Cai, Tao Gu, and Wei Shao. 2019. Your eyes reveal your secrets: An eye movement based password inference on smartphone. *IEEE transactions on mobile computing* 19, 11 (2019), 2714–2730.

[54] Zhen Xiao, Tao Chen, Yang Liu, and Zhenjiang Li. 2020. Mobile phones know your keystrokes through the sounds from finger's tapping on the screen. In *2020 IEEE 40th International Conference on Distributed Computing Systems (ICDCS)*. IEEE, 965–975.

[55] Xiaoqiang Xu, Xuanqi Meng, Xinyu Tong, Xiulong Liu, Xin Xie, and Wenyu Qu. 2024. HyperTracking: Exploring the hyperbolic model for non-line-of-sight device-free Wi-Fi tracking. *Proceedings of the ACM on Interactive, Mobile, Wearable and Ubiquitous Technologies* 7, 4 (2024), 1–26.

[56] Huan Yan, Jian Liu, Xiang Zhang, Zhi Liu, Bin Liu, Meng Li, Zheng Gong, Ming Gao, and Fusang Zhang. 2026. DifLoc+: Toward Robust Wi-Fi Hidden Camera Localization Based on Electromagnetic Difraction. *IEEE Journal on Selected Areas in Communications* 44 (2026), 4223–4238. doi:10.1109/JSAC.2026.3671737

[57] Huan Yan, Xiang Zhang, Jinyang Huang, Yuanhao Feng, Meng Li, Anzhi Wang, Weihua Ou, Hongbing Wang, and Zhi Liu. 2025. Wi-sfdagr: Wifi-based cross-domain gesture recognition via source-free domain adaptation. *IEEE Internet of Things Journal* (2025).

[58] Edwin Yang, Song Fang, Ian Markwood, Yao Liu, Shangqing Zhao, Zhuo Lu, and Haojin Zhu. 2022. Wireless training-free keystroke inference attack and defense. *IEEE/ACM Transactions on Networking* 30, 4 (2022), 1733–1748.

[59] Edwin Yang, Qiuye He, and Song Fang. 2022. WINK: Wireless inference of numerical keystrokes via zero-training spatiotemporal analysis. In *Proceedings of the 2022 ACM SIGSAC Conference on Computer and Communications Security*. 3033–3047.

[60] Zhuolin Yang, Yuxin Chen, Zain Sarwar, Hadleigh Schwartz, Ben Y Zhao, and Haitao Zheng. 2023. Towards a general video-based keystroke inference attack. In *32nd USENIX Security Symposium (USENIX Security 23)*. 141–158.

[61] Zhiyun Yao, Kai Niu, Xuanzhi Wang, Rong Zheng, Junzhe Wang, Duo Zhang, and Daqing Zhang. 2025. WiCaliper: Simultaneous Material and 3D Size Sensing for Everyday Objects Using WiFi. *IEEE Journal on Selected Areas in Communications* (2025).

[62] Enze Yi, Dan Wu, Jie Xiong, Fusang Zhang, Kai Niu, Wenwei Li, and Daqing Zhang. 2024. BFMSense:WiFi sensing using beamforming feedback matrix. In *21st USENIX Symposium on Networked Systems Design and Implementation (NSDI 24)*. 1697–1712.

[63] Qinggang Yue, Zhen Ling, Xinwen Fu, Benyuan Liu, Wei Yu, and Wei Zhao. 2014. My google glass sees your passwords. *Proceedings of the Black Hat USA* (2014).

[64] Daqing Zhang, Kai Niu, Jie Xiong, Fusang Zhang, and Xuanzhi Wang. 2023. WiFi/4G/5G Based Wireless Sensing: Theories, Applications and Future Directions. In *Integrated Sensing and Communications*. Springer, 387–417.

[65] Xiang Zhang, Yu Gu, Huan Yan, Yantong Wang, Mianxiong Dong, Kaoru Ota, Fuji Ren, and Yusheng Ji. 2023. Wital: A COTS WiFi devices based vital signs monitoring system using NLOS sensing model. *IEEE Transactions on Human – Machine Systems* 53, 3 (2023), 629–641.

[66] Xiang Zhang, Jinyang Huang, Huan Yan, Yuanhao Feng, Peng Zhao, Guohang Zhuang, Zhi Liu, and Bin Liu. 2025. WiOpen: A Robust Wi-Fi-Based Open-Set Gesture Recognition Framework. *IEEE Transactions on Human–Machine Systems* 55, 2 (2025), 234–245.

[67] Xiang Zhang, Huan Yan, Jinyang Huang, Bin Liu, Yuanhao Feng, Jianchun Liu, Meng Li, Fusang Zhang, and Zhi Liu. 2026. Beyond Physical Labels: Redefining Domains for Robust WiFi-based Gesture Recognition. *Proceedings of the ACM on Interactive, Mobile, Wearable and Ubiquitous Technologies* 10, 1 (2026), 1–27.

[68] Xiang Zhang, Jie Zhang, Zehua Ma, Jinyang Huang, Meng Li, Huan Yan, Peng Zhao, Zijian Zhang, Bin Liu, Qing Guo, Tianwei Zhang, and NengHai Yu. 2025. CamLoPA: A Hidden Wireless Camera Localization Framework via Signal Propagation Path Analysis. In *2025 IEEE Symposium on Security and Privacy (SP)*. Los Alamitos, CA, USA, 3653–3671.

[69] Xiang Zhang, Jie Zhang, Huan Yan, Jinyang Huang, Zehua Ma, Bin Liu, Meng Li, Kejiang Chen, Qing Guo, Tianwei Zhang, and Zhi Liu. 2025. DifLoc: WiFi Hidden Camera Localization Based on Electromagnetic Difraction. In *The 34th USENIX Security Symposium*. Seattle, WA, USA.

[70] Zijian Zhang, Nurilla Avazov, Jiamou Liu, Bakh Khoussainov, Xin Li, Keke Gai, and Liehuang Zhu. 2020. WiPOS: A POS terminal password inference system based on wireless signals. *IEEE Internet of Things Journal* 7, 8 (2020), 7506–7516.

[71] Man Zhou, Qian Wang, Jingxiao Yang, Qi Li, Feng Xiao, Zhibo Wang, and Xiaofeng Chen. 2018. Patternlistener: Cracking android pattern lock using acoustic signals. In *Proceedings of the 2018 ACM SIGSAC Conference on Computer and Communications Security*. 1775–1787.

[72] Tong Zhu, Qiang Ma, Shanfeng Zhang, and Yunhao Liu. 2014. Context-free